\documentclass{elsart}

\usepackage{amssymb}

\journal{Astroparticle Physics}

\begin{document}

\begin{frontmatter}



\title{New constraints on space-time Planck scale fluctuations
from established high energy astronomy observations}


\author{R. Le Gallou}
\ead{roland.le-gallou@durham.ac.uk}

\address{University of Durham, Department of Physics, Rochester Building, 
Science Laboratories, South Road, Durham DH1 3LE, UK}

\begin{abstract}
The space-time metric is widely believed to be subject to stochastic
fluctuations induced by quantum gravity at the Planck scale. This
work is based on two different phenomenological approaches being currently
made to this topic, and theoretical models which describe this phenomenon 
are not dealt with here. By using the idea developed in one of these two
approaches in the framework of the other one, it is shown that the
constraints on the nature of Planck scale space-time fluctuations
already set by the observation of electrons and gamma-rays with energies
above 15 TeV are much stronger than have been shown so far. It is
concluded that for the kind of Planck scale fluctuations implied by
several models, including the most naive one, to be consistent with
the observations, the transformation laws between different reference
frames must be modified in order to let the Planck scale be observer-independent. 
\end{abstract}

\end{frontmatter}

\section{Introduction}

Merging relativity and quantum mechanics is one of the greatest challenges
of modern physics. It is a very difficult task because the models
involved in it are hardly testable experimentally, if at all. Though,
recently, various attemps to do so have appeared in the literature.
This paper focuses on two particular approaches made in this respect, which
both search for modifications of kinematics and photon propagation
at high energy by adding an extra term to the dispersion relation.
These two approaches differ in the nature of this term, and hence 
obtain different kinds of results. The first approach \cite{Aloisio 1,Aloisio 2,Lieu & Hillman,Ng et al 2}
searches for effects induced by Planck scale space-time fluctuations,
therefore introducing a stochastic term in the dispersion relation,
whereas the second one \cite{JLM 1,JLM 2,GAC et al 1998,GAC and Piran,Gonzalez-Mestres}
introduces a constant term in the dispersion relation. Both approaches
have succeeded in constraining the parameter they add to the dispersion
relation using experimental data from high energy astronomy.
Other kinds of observations also constrain these parameters even more
strongly. In the framework of the first approach such constraints
come from optical interferometry \cite{Lieu & Hillman} but they are
controversial \cite{Ng et al comment,Coule comment}. In the case
of the second approach they come from atomic and nuclear physics experiments
\cite{Sudarsky et al}.

The aim of this paper is to show that the application to the first
approach of one of the developments made in the framework of the second
approach allows us to derive new constraints on the nature of space-time
at Planck scale. In this respect, the parametrization of the dispersion
relation modifying term will be generalized as being the sum of a
constant and a stochastic term. In the following paragraphs the first
approach will be introduced, and the modification of the dispersion
relation due to Planck scale fluctuations will be explained. The implications
of the latter concerning the nature of the laws of coordinate transformation
between different inertial reference frames will be discussed. Then,
one of the developments made in the framework of the second approach
will be introduced, applied to the first approach and some details considered.
The conclusion and a short discussion will then follow.

\section{First approach: the effects of Planck scale space-time fluctuations
on kinematics}

Let us first introduce space-time Planck scale fluctuations naively. Using
Heisenberg's uncertainty principle one can find that virtual particles
having the Planck energy can pop in and out of the vacuum within the
Planck scale of time and space. The spatial extension of such particles
would match the Schwarzschild radius associated to their mass, therefore
their presence would curve the space-time continuum into something
looking like a foam. This is where the concept of {}``space-time
foam'' comes from. This description is very naive, and quantum gravity
models, of course, introduce the phenomenon differently. An interesting
concept related to this topic is that of quanta of length and area
\cite{John Baez}. Meanwhile, space-time Planck scale fluctuations
have not yet been observed. Following the work of \cite{Aloisio 1,Aloisio 2},
we will introduce stochastic fluctuations into the space-time metric.

If space-time is subject to stochastic fluctuations at the Planck
scale, then each measurement of lengths and time intervals must be
affected by fluctuating terms and one could write:

\begin{equation}
\label{eq: l+-sigma l}
\left\{ \begin{array}{c}
l\rightarrow l\pm \sigma _{l}\; ;\; \sigma _{l}\simeq l_{P}\\
t\rightarrow t\pm \sigma _{t}\; ;\; \sigma _{t}\simeq t_{P}
\end{array}\right. 
\end{equation}
where \( l_{P} \) and \( t_{P} \) are the Planck length and time
intervals respectively. As developed in \cite{Aloisio 1,Aloisio 2},
assuming that the de Broglie wavelengths of particles follow the fluctuations
of space time (\emph{assumption} \emph{1}) leads to, using \( \hbar  \)=c=1,

\begin{equation}
\label{eq: dT dlambda}
\left\{ \begin{array}{c}
\delta T\simeq t_{P}\Rightarrow \delta E=\delta \left( \frac{1}{T}\right) 
\simeq E^{2}t_{P}=\frac{E^{2}}{E_{P}}\\
\delta \lambda \simeq l_{P}\Rightarrow \delta p=\delta \left( \frac{1}{\lambda }\right) 
\simeq p^{2}l_{P}=\frac{p^{2}}{E_{P}}
\end{array}\right. 
\end{equation}
 \( E_{P} \) being the Planck energy. Assuming that the fluctuations
on t and each component of l are uncorrelated (\emph{assumption} \emph{2}:
rotational invariance) , one can write \cite{Aloisio 1,Aloisio 2}:

\begin{equation}
\label{eq: <E> + alpha dE}
\left\{ \begin{array}{c}
E\simeq \overline{E}+\zeta \frac{\overline{E}^{2}}{E_{P}}\\
p\simeq \overline{p}+\xi \frac{\overline{p}^{2}}{E_{P}}
\end{array}\right. 
\end{equation}
with \( \zeta  \) and \( \xi  \) being distributed as a gaussian
of mean value \( \mu =0 \) and variance \( \sigma =1 \).

A more general way of describing the space-time fluctuations at the
Planck scale is to write \( \sigma _{x}/x=f(x_{P}/x) \), where \( x \)
stands for \( l \) or \( t \), with \( f\ll 1 \) for \( x\gg x_{P} \)
and \( f\gtrsim 1 \) for \( x\lesssim x_{P} \). In this case, \( f(x) \)
can be approximated with the lower order term of its expansion in
the range \( x\gg x_{P} \) in the following way \cite{Lieu & Hillman}: 

\begin{equation}
\label{eq: sigmat/t first order term}
\frac{\sigma _{x}}{x}\simeq a_{0}\left( \frac{x_{P}}{x}\right) ^{\alpha }
\end{equation}
where both \( \alpha  \) and \( a_{0} \) are positive constants
of order 1. The naive choice for \( \alpha  \) is 1, which is equivalent
to eq.\ref{eq: l+-sigma l} and is indeed the first order term given
by quantum loop gravity (see \cite{Lieu & Hillman}), but it is worth
keeping in mind that some models of quantum space-time give other
values like \( \alpha =1/2 \) (random-walk scenario) or \( \alpha =2/3 \)
(holographic principle of Wheeler and Hawking). A discussion about
this and citations of these models are given in \cite{Lieu & Hillman,Ng et al 2}
and references therein.\\

The choice of the reference frame in which to apply the above equations
raises an important issue concerning special relativity: in which
reference frames do the fluctuations have the Planck scale, if they
exist? Consider the three following cases:

\begin{itemize}
\item \textbf{case A:} Planck scale space-time fluctuations do not exist.
\item \textbf{case B:} If the fluctuations have the Planck scale in all reference
frames, then the laws of coordinate transformations between different
inertial reference frames would have to depart from pure Lorentz transformations
to let this scale be invariant. This is the milestone of Doubly Special
Relativity (DSR) theories, in which both the velocity of light and
the Planck scale of length and mass are observer-independant scales
\cite{Amelino-Camelia 1,Magueijo & Smolin,Amelino-Camelia 2,Amelino-Camelia 3,Amelino-Camelia 5}.
\item \textbf{case C:} Space-time fluctuations may have the Planck
scale in one preferred reference frame only, and boosted values of
this scale in other reference frames. Indeed, there is a preferred
reference frame in the Universe: the one where the Cosmic Microwave
Background (CMB) appears isotropic. This case has been considered
in many phenomenological studies of the effects that a fluctuating
space-time would have on kinematics \cite{JLM 1}, although it implies
the abandonment of the relativity principle which stipulates that
laws of physics should be the same for all inertial observers.
\end{itemize}
Expanding eq.\ref{eq: <E> + alpha dE} at the first order in \( E/E_{P} \)
in the usual dispersion relation, in the energy range \( m^{2}\ll E^{2}\simeq p^{2}\ll E_{P}^{2} \)
and using assumption 2, one obtains:\begin{equation}
\label{eq: dispersion relation}
m^{2}=E^{2}-p^{2}+\eta \frac{E^{3}}{E_{P}}
\end{equation}
 where \( \eta  \) is distributed as a gaussian with \( \mu =0 \)
and \( \sigma =2\sqrt{2} \). This expression is obtained in \cite{Aloisio 1,Aloisio 2}
as well, although the value of \( \eta  \)'s distribution width is
only assumed to be of order unity in the latter. However the \( 2\sqrt{2} \)
value used here can be retrieved in \cite{Lieu & Hillman}. One can
notice that this expression is not invariant under ordinary Lorentz
transformations: it is either valid in one reference frame only, or
valid in all possible reference frames if the laws of coordinates
transformation between different inertial reference frames were to
depart from ordinary Lorentz transformations.

Using equation \ref{eq: sigmat/t first order term} instead of \ref{eq: <E> + alpha dE}
as a starting point, one can derive a more general expression for
the dispersion relation: \begin{equation}
\label{eq: general dispersion relation}
m^{2}=E^{2}-p^{2}+\frac{\eta E^{\alpha +2}}{E_{P}^{\alpha }}
\end{equation}
 equation \ref{eq: dispersion relation} being the case \( \alpha =1 \)
of this equation. 

Equations \ref{eq: dispersion relation} and \ref{eq: general dispersion relation}
express the fact that at each measurement, the measured values of
E and p are different from their mean value. Consider interactions
where the energy exchanged by the particles involved is \( \ll E_{P} \).
The typical scales of length and time of these interactions are much
larger that the Planck ones. Hence, one should stipulate independent
fluctuations for each initial and final particle \cite{Aloisio 2}.
It is worth noticing that the conservation of energy-momentum then
still applies on the macroscopic scale for the mean values of E and
p, but does not apply any more in individual interactions within the
amplitude defined by the fluctuating term of equations \ref{eq: dispersion relation}
and \ref{eq: general dispersion relation}.

\section{Introduction of the second approach and demonstration}

In the analysis performed in \cite{Aloisio 1,Aloisio 2} and \cite{Lieu & Hillman}
\( \eta  \) is a stochastic variable which accounts for the space-time
fluctuations specified in eq.\ref{eq: l+-sigma l}. On the other hand,
it is introduced as a constant in the analysis performed in \cite{JLM 1,JLM 2},
\cite{GAC et al 1998,GAC and Piran} and \cite{Gonzalez-Mestres}. In these approaches
the modification of the dispersion relation does not come explicitly
from space-time fluctuations. Instead, it comes from the introduction
of a minimum length in \cite{GAC et al 1998,GAC and Piran} (see \cite{Magueijo & Smolin,Amelino-Camelia 2})
and \cite{Gonzalez-Mestres}, or from quantum field theory or theoretical
approaches to quantum gravity in \cite{JLM 1,JLM 2}. In the latter
approach, \( \eta  \) depends on the nature of the particle involved
and does not have to be of order 1. The general form of the modified
dispersion relation is then:\begin{equation}
\label{eq: modified dispersion relation with eta fixed}
m_{k}^{2}=E^{2}_{k}-p_{k}^{2}+\frac{\eta _{k}E_{k}^{\alpha +2}}{M_{0}^{\alpha }}
\end{equation}
 where \( M_{0} \) is of the order of \( E_{P} \), and the subscript
\( k \) stands for the nature of the particle involved.\\

A priori the dispersion relation could also be altered by both a constant
term and a stochastic term due to Planck scale space-time fluctuations.
Let's write \( \eta =\eta _{a}+\eta _{s} \), where \( \eta _{a} \)
is the constant term and \( \eta _{s} \) the stochastic one. From
now on, the constant and the stochastic \( \eta  \)
approaches will be referred to as the \( \eta =\eta _{a} \) and \( \eta =\eta _{s} \)
approaches respectively. 

Many phenomenological studies of Lorentz violating kinematics have
been made following either the \( \eta =\eta _{a} \) or \( \eta =\eta _{s} \)
approach. Consider here the case where \( \eta =\eta _{a}+\eta _{s} \).
The key point of this paper is to show that conclusions can be obtained
by using a study already done in the \( \eta =\eta _{a} \) case and
bringing it into the case of \( \eta =\eta _{a}+\eta _{s} \). This
study concerns the 1 vertex electromagnetic interactions involving
an electron and a photon, which are forbidden in usual kinematics
by the conservation laws of E and p but can be allowed if Planck scale
space-time fluctuations are taken into account.

Equation \ref{eq: dispersion relation} shows that in electromagnetic
interactions involving electrons the fluctuating term becomes significant
when \( E\geq E_{C}= \) \( \left( E_{P}m^{2}\right) ^{1/3}\simeq  \)15
TeV, where \( m \) is the mass of the electron. If we consider case
C, this would hold in the CMB reference frame only and the TeV photons
above 15 TeV such as those seen by CANGAROO \cite{Tanimori et al}
and HEGRA \cite{Aharonian et al} would obey modified kinematics,
as would the very high energy (~100 TeV) electrons inferred from ASCA X-ray
observations of the Crab nebula \cite{Koyama et al}. On the other hand, 
if the Universe is in case B then space-time fluctuations have
the same scale in all reference frames, hence one can choose the centre
of mass of the interaction to determine \( E_{C} \). Modified kinematics
would then occur in interactions where particles have \( E\geq E_{C} \)
in the centre of mass. In this case the Planck scale fluctuation term
of eq.\ref{eq: dispersion relation} would be negligible in the above
processes since the energy they involve in their centre of mass is
\( \ll E_{C} \). 

In \cite{JLM 1,JLM 2} and \cite{Amelino-Camelia 4}, it has been
shown that the above observations of very high energy photons and
electrons rule out the possibility that \( \eta \simeq -1 \) in the
case \( \eta =\eta _{a} \) and \( \alpha =1 \), in the framework
of case C. The demonstration is based on the following argument: if
\( \eta <0 \) in case C, photons and electrons of energy above
\( E_{C} \) can undergo 1 vertex interactions. As a result the photons
and electrons of energies above \( E_{C} \) would respectively decay
into \( e^{+}e^{-} \) pairs and radiate spontaneously, and hence
would not be observed. On the other hand, the case \( \eta \simeq +1 \)
is still allowed in the case \( \eta =\eta _{a} \), because it does not
allow the 1 vertex interactions to happen. These conclusions are the
same for any value of \( \alpha \leq 1 \).

Now, what happens to these conclusions if \( \eta =\eta _{a}+\eta _{s} \),
if we assume \( \left| \eta _{a}\right| \lesssim 1 \) 
(\emph{assumption} \emph{3}) ? In this case \( \eta  \) could take
any sign, hence 1 vertex interactions would be made possible for any
value of \( \alpha \leq 1 \). Since such interactions have not been
observed, it means that case C would be ruled out if the assumptions
made so far are correct. This would imply that Planck scale space-time
transformations don't exist (case A) or that Lorentz transformations
would have to be rewritten as it is aimed at in DSR theories (case
B). If on the other hand one or more of the assumptions made so far
was wrong, it would be a very important clue for the development
of quantum gravity theories. Experimental
results in atomic and nuclear physics have already set the constraint
\( \left| \eta _{a}\right| \ll 1 \) in the case \( \alpha =1 \)
\cite{Sudarsky et al}, which confirms assumption 3. 
(High energy astronomy also provides constraints on \( \left| \eta _{a}\right| \) 
\cite{Schaefer}, but they are not as stringent as those from \cite{Sudarsky et al}.)\\

Let us consider in more detail how these assertions can be demonstrated
by calculating the cosine of the angle \( \theta  \) between the
two outgoing particles of the 1 vertex interactions in the case \( \alpha =1 \)
\cite{Amelino-Camelia 4}. Let us assume here that the standard conservation
of E and p still applies in the DSR framework. This is not the case
according to the developments made in \cite{Amelino-Camelia 2}. However,
the modification of this conservation law would not be a fluctuating
term, but would come from the effects of the constant component of
the modified dispersion relation which has been severely constrained
by \cite{Sudarsky et al}. Hence this assumption should not affect
the result qualitatively.

Using the unmodified dispersion relations, one finds respectively,
in the processes \( \gamma \rightarrow e^{+}e^{-} \)and \( e^{\pm }\rightarrow e^{\pm }\gamma  \)
: \begin{equation}
\label{eq: cos theta unmodified}
\left\{ \begin{array}{c}
\cos \theta =1+\frac{m^{2}}{2E^{2}}\frac{1}{u^{2}(1-u)^{2}}>1\\
\cos \theta =1+\frac{m^{2}}{2E^{2}}\frac{1}{(1-u)^{2}}>1
\end{array}\right. 
\end{equation}
where \( E \) is the energy of the initial particle and \( u \)
is the fraction of it taken by the positron in the first line and
by the photon in the second line. In both cases the cosine is \( >1 \) and
thus unphysical: the interactions are forbidden. 

Now, if one uses the modified dispersion relation and a first order
expansion in \( \left( m/E\right) ^{2} \)\( \sim E/E_{P}\sim 10^{-15} \),
one obtains respectively:

\begin{equation}
\label{eq: cos theta modified}
\left\{ \begin{array}{c}
\cos \theta =1+\frac{m^{2}}{2E^{2}}\frac{1}{u^{2}(1-u)^{2}}+\frac{E}{2E_{P}}
\left[ \eta _{p}u\left( 1-\frac{u}{1-u}\right) +\eta _{e}(1-u)\left( 1-\frac{1-u}{u}\right) 
+\eta _{\gamma \gamma }\frac{1}{u(1-u)}\right] \\
\cos \theta =1+\frac{m^{2}}{2E^{2}}\frac{1}{(1-u)^{2}}+\frac{E}{2E_{P}}
\left[ \eta _{i}\frac{1}{u(1-u)}-\eta _{f}\left( 1-u+\frac{(1-u)^{2}}{u}\right)
 -\eta _{\gamma \gamma }\left( u+\frac{u^{2}}{1-u}\right) \right] 
\end{array}\right. 
\end{equation}
 where in the first line \( \eta _{\gamma \gamma },\eta _{e},\eta _{p} \)
apply respectively to the photon, the electron and the positron, and
in the second line \( \eta _{i},\eta _{f},\eta _{\gamma \gamma } \)
apply respectively to the initial electron, the final electron and
the photon.

Both cosines can be physical if, respectively:\begin{equation}
\label{eq: condition on u}
\left\{ \begin{array}{c}
\eta _{p}u^{3}(1-u)(1-2u)+\eta _{e}u(1-u)^{3}(2u-1)+\eta _{\gamma \gamma }u(1-u)
<-\left( \frac{E_{C}}{E}\right) ^{3}\\
\eta _{i}\frac{1-u}{u}-\eta _{f}\frac{(1-u)^{3}}{u}-\eta _{\gamma \gamma }u(1-u)
<-\left( \frac{E_{C}}{E}\right) ^{3}
\end{array}\right. 
\end{equation}
Since the various \( \eta  \) terms can take stochastic values, these
conditions are allowed. In both relations the leading \( \eta  \)
term is the one multiplied by the lowest power of \( u \) or \( (1-u) \),
which is the one associated with the initial particle. It is interesting
to note that the second relation has an IR divergence, similar
to its 2 vertex equivalent : if u is allowed to be \( \ll 1 \), this
relation can be satisfied even when \( E\ll E_{C} \). This divergence
doesn't have to be treated here, since the study of the phenomenon
is conclusive at high energy anyway.

\section{Conclusion}

The conclusion of this paper is that Planck scale space-time fluctuations
described by an exponent \( \alpha \leq 1 \) are consistent with
the observations only if the Planck scale is observer-independent,
in the framework of the assumptions made here. Concerned models are:
the naive description of space-time, as described by eq.\ref{eq: l+-sigma l}
\cite{Aloisio 1,Aloisio 2} and implied by quantum loop gravity; the
random walk scenario; and the holographic principle of Wheeler and
Hawking (see \cite{Lieu & Hillman}). As a result, there are two possibilites.
The first one is that Planck scale space-time fluctuations don't exist.
The second one is that if they exist according to one of the above
models the Planck scale has to be observer-independent, which implies
that the laws of coordinate transformations between different inertial
reference frames have to be changed. If both
possibilites were ruled out by other means, one could show that at
least one of the assumptions made here is wrong, or that \( \alpha >1 \).
This would also be a significant clue concerning the development
of quantum gravity theories. \\

\emph{Remarks:} A similar conclusion has been reached very recently in \cite{Aloisio 3}, a paper which was 
released on the astro-ph web site two weeks before the submission of the present paper.
Reference \cite{Lieu & Hillman} is a recent study of Planck scale
space-time fluctuations based on stellar interferometry. It is cited
several times in this paper, regarding the developments made in its
beginning. It concludes that space-time does not fluctuate at the Planck scale;
however, the demonstration it uses is controversial \cite{Ng et al comment,Coule comment}.
Laser interferometers which will be used in gravitational wave detectors
like VIRGO and LIGO will also be able to detect Planck scale space-time
fluctuations if they exist, or rule them out \cite{Amelino-Camelia 6},
giving one more independent insight on the nature of space-time at
the Planck scale.

\ack I wish to thank particularly P.M.Chadwick, J.McKenny and J.L.Osborne for their useful comments
about the writing of this paper. The support of PPARC through the research grant 
PPA/G/S/2000/00033 is gratefully acknowledged.



\end{document}